\documentclass[prl,twocolumn,showpacs]{revtex4}
\usepackage{graphicx}
\newcommand{\n}{\nonumber}
\newcommand{\bn}{\begin{eqnarray}}
\newcommand{\en}{\end{eqnarray}}
\newcommand{\h}{\hspace}
\def\Journal#1#2#3#4{{#1} {\bf #2}, #3 (#4)}
\def\PR{Phys. Rev.}
\def\PRL{Phys. Rev. Lett.}
\def\PRA{Phys. Rev. A}
\def\JMP{J. Math. Phys.}
\begin{document}
\title {Crossover from one to three dimensions for a gas of hard-core bosons}
\author{Kunal K. Das}
\email{kdas@optics.arizona.edu}
\author{M.D. Girardeau}
\email{girardeau@optics.arizona.edu}
\author{E.M. Wright}
\email{Ewan.Wright@optics.arizona.edu}
 \affiliation{Optical
Sciences Center and Department of Physics, University of Arizona,
Tucson, AZ 85721}
\date{\today}
\begin{abstract}
We develop a variational theory of the crossover from the
one-dimensional (1D) regime to the 3D regime for ultra-cold Bose
gases in thin waveguides. Within the 1D regime we map out the
parameter space for fermionization, which may span the full 1D
regime for suitable transverse confinement.
\end{abstract}
\maketitle
It has recently been proved by Lieb and Seiringer \cite{LS} that
in a suitably-defined dilute limit, the many-body ground state of
a trapped ultra-cold gas of bosons in two or three dimensions
exhibits Bose-Einstein condensation into that orbital which
minimizes the Gross-Pitaevskii (GP) energy functional, and in fact
that the condensation is \emph{complete} in the sense that the
condensed fraction is unity. Their work should be consulted for
precise definitions and hypotheses required for the proof; here we
reiterate only a few points relevant here. In the 3D case the
dilute limit is defined as $a\to 0$ and $N\to\infty$ with both the
trap potential and $Na$ fixed where $a$ is the s-wave scattering
length and $N$ the number of particles, and in the 2D case it is
defined as $a\to 0$ and $N\to\infty$ with the trap potential and
$N/|\ln(a^{2}N)|$ fixed. They point out that in 1D their proof
fails, where there is presumably no Bose-Einstein condensate (BEC)
even at zero temperature (many-body ground state). In fact, the
exact many-body ground state of a \emph{spatially uniform}
(untrapped) 1D Bose gas with repulsive zero-range (delta function)
interaction was found long ago by Lieb and Liniger (LL) \cite{LL}
for all values of a 1D coupling constant $g_{1D}$, and was shown
by them to reduce for $g_{1D}\to\infty$ to the exact many-body
ground state of the impenetrable point Bose gas (``Tonks gas'')
found previously by one of us \cite{map}, for which Lenard proved
rigorously \cite{Lenard} that the occupation of the lowest orbital
is bounded above by $const.\sqrt{N}$, ruling out BEC (occupation
proportional to $N$). In the case of a \emph{trapped} Tonks gas no
such rigorous bound is known, but our numerical evaluation of the
largest eigenvalue of the reduced single-particle density matrix
of our exact many-body ground state \cite{GW3} suggests strongly
that the most highly occupied orbital has occupation behaving like
$N^p$ with $0<p<1$, again indicating absence of true BEC. In what
follows we will use the term 1D condensate to describe the
ground state of the trapped gas when the the system is still 1D,
in that it has the transverse profile of the trap ground state,
but the energy has deviated below that for an impenetrable Tonks
gas.

It is clear from the above discussion that for real atom
waveguides, for which the idealized limits $a\to 0$ and
$N\to\infty$ do not strictly apply, a crossover must occur from a
effectively 1D system, applicable when the waveguide is so long
and narrow (high transverse frequency) that transverse excitations
are frozen, to a 3D system with BEC accurately treated by the GP
equation (weaker transverse binding), the basis of most
theoretical work on trapped BECs. Detailed analysis of this
crossover is important for comparison with experiments, since the
1D regime has already been achieved experimentally
\cite{DePMcCWin99,BonBurDet01,GorVogLea01,Greiner} and the Tonks
regime (1D \emph{and} sufficiently large $g_{1D}$) is being
approached \cite{Greiner}. The dynamical reduction from 3D to 1D
and precise conditions on parameters necessary for achievement of
both the 1D limit and the Tonks-gas limit of the 1D regime have
been discussed in detail by Olshanii \cite{Olshanii} and by Petrov
\textit{et al.} \cite{Petrov}. In addition, Dunjko {\it et al.}
\cite{crossover2} have investigated the crossover between the
Thomas-Fermi and Tonks-Girardeau regimes in a 1D trap. Here we
note only that the 1D regime occurs when the waveguide is so thin
(transverse frequency so high) and density and temperature so low
that the longitudinal thermal and zero-point energies are both low
compared with the lowest transverse excitation energy, resulting
in ``freezing out'' all transverse excitations. Achievement of the
Tonks limit requires, in addition, that the scattering length $a$
is large enough and/or 1D density $n$ low enough that
$\hbar^{2}n/mg_{1D}\ll 1$ where the effective 1D coupling constant
is $g_{1D}=2\hbar^{2}a/m\ell_{0}^{2}$ and
$\ell_{0}=\sqrt{\hbar/m\omega_{0}}$ is the transverse oscillator
length. \vspace{-0.1cm}

\textit{Trap geometry}: A particularly convenient
geometry for discussing the crossover is a toroidal trap of high
aspect ratio $R=L/\ell_{0}$. The transverse trap potential is
symmetric about an axis consisting of a circle on which the trap
potential is minimum, and harmonic with respect to a coordinate
$\rho$ measured transversely with respect to this circle. This
geometry can equally well be interpreted as an infinitely long,
straight cylindrical waveguide with periodic boundary conditions
in the longitudinal direction. Toroidal traps of this form have
been experimentally produced and loaded \cite{WAD,Sauer}.

\textit{Hamiltonian}: We use a many-body Hamiltonian with
harmonic transverse binding and the usual Fermi pseudopotential interaction
$v({\bf r}_{ij})=4\pi a\delta({\bf r}_{ij})$ with a positive
s-wave scattering length $a$.  This leads to a well-defined
problem in 1D, the LL model \cite{LL}. Our toroidal system is
``almost 1D'' since the
transverse dimensions are confined, and we find that the
variational problem with the Fermi pseudopotential does not
encounter the difficulties (divergences and poorly-posed
variational problem) \cite{Huang1,Huang2} found in the 3D case.
The Hamiltonian is then
\begin{equation}
\hat{H}=\sum_{j=1}^{N}\left(-\frac{\hbar^{2}}{2m}\nabla_{j}^{2}
+\frac{1}{2}m\omega_{0}^{2}\rho_{j}^{2}\right)
+g_{3D}\h{-3mm}\sum_{1\le j<\ell\le N}\delta({\bf r}_{j}-{\bf
r}_{\ell})
\end{equation}
where $g_{3D}=4\pi a\hbar^{2}/m$ is the 3D coupling constant. The
Laplacian is to be expressed in cylindrical coordinates ${\bf
r}_{j}=(z_{j},\rho_{j},\theta_{j})$ where $z_j$, with $0\le
z_{j}\le L$, is a 1D coordinate measured around the torus
circumference, $\rho_j$ is a transverse radial coordinate measured
from the central circular torus axis, and $\theta_j$ is the
azimuthal angle about this axis.

\textit{Variational ground state}: We use a trial variational
ground state which assumes factorization of longitudinal and
transverse parts, with the transverse part depending on a single
transverse orbital $\phi_{tr}$ independent of azimuthal angle:
\begin{equation}
\Phi_{0}({\bf r}_{1},\cdots,{\bf
r}_{N})=\Phi_{long}(z_{1},\cdots,z_{N})
\prod_{j=1}^{N}\phi_{tr}(\rho_{j}) \quad .
\end{equation}
Use of a single transverse orbital is justified in two different
limits: (a) tight transverse confinement, transverse excitations
frozen, $\phi_{tr}$ is the unperturbed transverse oscillator
ground state; (b) weak transverse confinement and low density,
$\phi_{tr}$ is the transverse part of GP orbital, assuming
factorization of longitudinal and transverse parts of this GP
orbital. Note, in connection with case (b), that a GP orbital well
approximated by a Gaussian with respect to both $z$ and $\rho$
factorizes automatically, as does any Gaussian. More generally, we
leave the functional form of $\phi_{tr}$ free, to be determined as
part of the minimization of the variational ground state energy
$E_{0}=\langle\Phi_{0}|\hat{H}|\Phi_{0}\rangle$. Assuming
$\phi_{tr}$ and $\Phi_{long}$ normalized according to
\begin{eqnarray}
& &\int_{0}^{\infty}2\pi\rho d\rho\ |\phi_{tr}(\rho)|^{2}=1 \nonumber \\
& &\int_{0}^{L}dz_{1}\cdots\int_{0}^{L}dz_{N}
|\Phi_{long}(z_{1},\cdots,z_{N})|^{2}=1
\end{eqnarray}
one finds that the energy expectation value of $\Phi_0$ decomposes
as $E_{0}=N\epsilon_{tr}+E_{long}$ with
\begin{eqnarray}
\epsilon_{tr}&=&\int_{0}^{\infty}\h{-2mm}2\pi\rho d\rho\
\phi_{tr}^{*}
\left(\frac{-\hbar^{2}}{2m\rho}\frac{\partial}{\partial\rho}\rho
\frac{\partial}{\partial\rho}+\frac{1}{2}m\omega_{0}^{2}\rho^{2}\right)
\phi_{tr} \nonumber \\
E_{long}&=&\int_{0}^{L}dz_{1}\cdots\int_{0}^{L}dz_{N}
\Phi_{long}^{*}\hat{H}_{long}\Phi_{long}
\end{eqnarray}
where $\hat{H}_{long}$ is an effective longitudinal Hamiltonian
\begin{equation}
\hat{H}_{long}=\sum_{j=1}^{N}\frac{-\hbar^{2}}{2m\rho}
\frac{\partial^{2}}{\partial z_{j}^{2}}+g_{1D} \sum_{1\le
j<\ell\le N}\delta(z_{j}-z_{\ell})
\end{equation}
and $g_{1D}$ is an effective 1D coupling constant
\begin{equation}
g_{1D}[\phi_{tr}]=g_{3D}\int_{0}^{\infty}2\pi\rho d\rho\
|\phi_{tr}(\rho)|^{4} \quad .
\end{equation}
Note that the separation of transverse and longitudinal energies
is only partial, since $g_{1D}$ depends on $\phi_{tr}$.  We take
the value of the $g_{1D}$ corresponding to the transverse ground
state, with oscillator length $\ell_{0}$, as a reference
\bn
 g=g_{1D}\left[\frac{1}{\sqrt{\pi}\
 \ell_{0}}e^{-\rho^{2}/2\ell_{0}^{2}}\right]
 =\frac{g_{3D}}{2\pi\ell_{0}^{2}}\en
 and define the fractional 1D coupling constant
 $\bar{g}_{1D}=g_{1D}/g$.  For a 1D system $\bar{g}_{1D}=1$
 whereas it decreases as the system crosses over to 3D.  The relevant
dimensionless intensive variable of a 1D system
$\gamma=(mg_{1D})/(\hbar^{2}n)$ \cite{LL} suggests a dimensionless
measure $\bar{n}=(\hbar^{2}n)/(mg)$ of the linear density $n=N/L$.
We should keep in mind that $g\propto \omega_{0}$ and will vary as
the transverse confinement changes.

The total energy $E_{0}$ is to be minimized with respect to
variation of both $\phi_{tr}$ and $\Phi_{long}$ subject to the
normalization constraints. We imagine this done in two steps,
first holding $\phi_{tr}$ constant and minimizing $E_{long}$ with
respect to $\Phi_{long}(z_{1},\cdots,z_{N})$, then minimizing the
resultant $E_{0}$ with respect to $\phi_{tr}(\rho)$. For fixed
$\phi_{tr}$, hence fixed $g_{1D}$, the global minimum of
$E_{long}$ is realized by the exact ground state of
$\hat{H}_{long}$, which is the well-known LL Bethe Ansatz solution
\cite{LL}. One may instead use some simpler variational trial
state for $\Phi_{long}$, obtaining an upper bound to the
longitudinal energy $E_{long}=N\epsilon_{long}$. This has then to
be minimized with respect to variation of $\phi_{tr}$ subject to
the normalization constraint:
\begin{equation}
\frac{\delta\epsilon_{tr}}{\delta\phi_{tr}^{*}(\rho)}
+\frac{\partial\epsilon_{long}}{\partial g_{1D}}
\frac{\delta g_{1D}}{\delta\phi_{tr}^{*}(\rho)}
-\mu_{tr}2\pi\rho\phi_{tr}(\rho)=0
\end{equation}
where the transverse chemical potential $\mu_{tr}$ is the Lagrange
multiplier for the transverse normalization constraint. Evaluation
of the functional derivatives leads to the following generalized
transverse GP equation:
\begin{eqnarray}
\mu_{tr}\phi_{tr} &=& -\frac{\hbar^{2}}{2m}\left(
\frac{\partial^2}{\partial\rho^2} +
\frac{1}{\rho}\frac{\partial}{\partial\rho} \right)\phi_{tr}
+\frac{1}{2}m\omega_{0}^2\rho^2\phi_{tr} \nonumber \\
&+& 2g_{3D}\frac{\partial\epsilon_{long}}{\partial g_{1D}}
|\phi_{tr}|^2\phi_{tr} ,
\end{eqnarray}
where $\mu_{tr}$ is to be adjusted so that $\phi_{tr}$ satisfies
the normalization constraint. The solution depends on $g_{1D}$
which in turn depends on $\phi_{tr}$, so the solution has to be
determined by a self-consistent iterative procedure, by making an
initial guess for $g_{1D}$, evaluating
$\partial\epsilon_{long}/\partial g_{1D}$ at this value of
$g_{1D}$, solving Eq. (8) for $\phi_{tr}$, determining a new
$g_{1D}$ from Eq. (6), and iterating to convergence.

Two of us have previously developed a variational theory
\cite{crossover1} assuming the same toroidal geometry and using a
variational trial state for $\Phi_{long}$ based on the variational
pair theory of many-boson systems \cite{GA}. In that case the
transverse GP equation (8) reduces to the previous one, Eq. (16)
of \cite{crossover1}, after correction of a typo therein
\cite{typo}. That work was devoted to investigation of the
BEC-Tonks crossover in the 1D regime where the transverse orbital
is frozen in the unperturbed transverse oscillator state, and no
transverse GP equation was solved. Here we are concerned with a
different crossover, namely the 1D-3D crossover, which depends
crucially on the solution of the transverse GP equation. In the
following two sections we shall separately consider first the case
where $\Phi_{long}$ is approximated by the GA variational pair
theory \cite{GA} as in \cite{crossover1}, which gives accurate
results except in the Tonks gas regime (1D limit \emph{and} very
large scattering length). Then we shall work out the solution
using the LL theory \cite{LL}, the exact ground state of
$\hat{H}_{long}$. This latter is accurate even in the Tonks
regime, but is more complicated to work out since the LL energy
per particle $\epsilon_{LL}$ is known only from numerical
solutions of the nonlinear LL integral equation.

\textit{Pair theory solution}: Approximating the longitudinal
energy by the 1D GA pair theory energy, we will first set
$\epsilon_{long}=\epsilon_{P}$, the expression for which we
derived in Ref.~\cite{crossover1} and can be written as
\bn\label{pairen}
\frac{\hbar^{2}}{m}\frac{\epsilon_{P}}{g^{2}}=\bar{g}_{1D}^{2}\left[\frac{1}{2\gamma}
-\frac{1}{3\pi}\sqrt{\frac{f^{3}}{\gamma}}\left[
(2-\lambda)E-\lambda
K\right]\right.\n\\\left.+\frac{f}{8\pi^{2}}\left[(1-\lambda)^{2}K^{2}
+[(1+\lambda)K-2E]^{2}\right]\right].\en
Here $K(1-\lambda)$ and $E(1-\lambda)$ are complete elliptic
integrals \cite{abramowitz} and $f$ is the Bose-condensed fraction
related to the dimensionless pair theory parameter $\lambda$
through the coupled equations
\bn \lambda=\sqrt{\frac{\gamma}{f}}\frac{[(1\!-\!\lambda)K]}{2\pi}
;\h{2mm}
  \frac{1-f}{f}=\sqrt{\frac{\gamma}{f}}\frac{[(1\!+\!\lambda)K\!-\!2E]}{2\pi} .\en
The partial derivative of the pair theory energy has to be taken
\emph{before minimization} \cite{crossover1} rather than that of
the expression in Eq.~(\ref{pairen}) which is true only at the
minimum
 \bn \frac{\partial{\epsilon_{P}}}{{\partial
g_{1D}}}=\frac{n}{2}\left[2-f^{2}
 -2\lambda f^{2}+\lambda^{2}f^{2}\right].
\en
\begin{figure}\vspace{-1cm}
\includegraphics*[width=\columnwidth,angle=0]{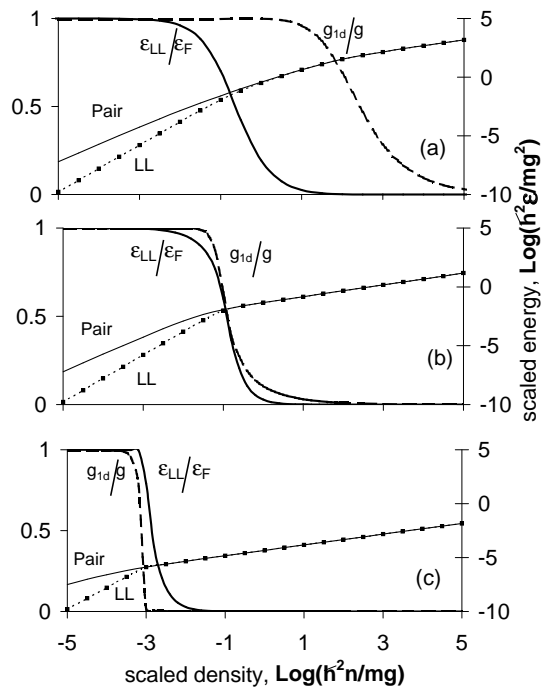}\vspace{-1cm}
\caption{Left axis: ratios
$\epsilon_{LL}/\epsilon_{F}$ and $g_{1D}/g$. Horizontal axis: logarithm of
scaled linear density $\bar{n}=(\hbar^{2}n)/(mg)$. Right axis: logarithms of
scaled energies calculated according to pair theory
$\hbar^{2}\epsilon_{P}/(mg^{2})$ and Lieb-Liniger theory
$\hbar^{2}\epsilon_{LL}/(mg^{2})$. The plots are shown for three different
strengths of the transverse trap frequencies $\omega_{0}$ directly
proportional to the value of the dimensionless quantity
$mg^{2}/(\hbar^{3}\omega_{0})=$ (a) $10^{-2}$, (b) $10^2$ and (c)
$10^8$. Note that the scale of actual linear densities $n$ differ
in the three plots since $g$ is different for each plot.}
\label{Fig1}\vspace{-6mm}
\end{figure}

\textit{LL solution}: Next we shall use the LL energy
$\epsilon_{LL}$ \cite{LL}, which is the global minimum of
$\epsilon_{long}$ with respect to unrestricted functional
variation of $\Phi_{long}$. It yields, together with Eq. (8), the
best possible variational solution obtainable with a trial state
of form (2):
\bn \frac{\hbar^2}{m}\frac{\epsilon_{LL}}{g^{2}}
=\frac{\bar{g}_{1D}^{2}}{2\gamma^{2}}e(\gamma),\en
where $e(\gamma)$ is obtained by solving the Lieb-Liniger system
of integral equations \cite{LL}. The partial derivative
 \bn
\frac{\partial{\epsilon_{LL}}}{{\partial
g_{1D}}}=\frac{\hbar^2}{2m}n^{2}e'(\gamma)\frac{m}{\hbar^{2}n}=
\frac{n}{2}e'(\gamma)\en
is the same regardless of the order in which the
expectation and the derivative are evaluated, unlike in pair
theory since the LL solution is an exact eigenstate of
$\hat{H}_{long}$ and thus obeys the Hellmann-Feynman theorem.

\textit{Numerical results and discussion}: For both pair theory
and Lieb-Liniger theory, the transverse GP equation was solved
numerically using the method of discrete variable representation
\cite{BayeHeenan}, a method based on Gauss quadrature with a
well-defined discrete normalization. Iterating the solution of the
GP equation alternately with the evaluation of the longitudinal
energies lead to self-consistent values for $g_{1D}$ and
$\epsilon_{long}$ in both theories. In using LL theory we used the
tabulated values of $e(\gamma)$ in \cite{OlshaniiData} for
intermediate values of $\gamma$ and the limiting expressions in
\cite{LL} for high and low values of $\gamma$.  For low $\gamma$,
i.e. for high linear densities, the energies in both pair theory
and LL theory coincide with the Bogoliubov energy $ng_{1D}/2$.

In the Tonks limit of completely impenetrable bosons, all of the
atoms are ``fermionized" and the energy per atom is simply the
energy of free fermions
\bn \frac{\hbar^2}{m}\frac{\epsilon_{F}}{g^{2}}
=\frac{\pi^{2}\bar{n}^{2}}{6}.\en
We can get a sense of the degree of ``fermionization" of the
system by evaluating the ratio $\epsilon_{LL}/\epsilon_{F}$. In
the Tonks limit this would be unity, but as the atoms become
penetrable the energy approaches a linear dependence on the
density so that the ratio will approach zero. In Fig.~\ref{Fig1}
we plot this ratio (for LL theory) as a function of the density
$\bar{n}$ for different values of the dimensionless quantity
$mg^{2}/(\hbar^{3}\omega_{0})$ proportional to the transverse
trapping frequency $\omega_{0}$. In the same figure we also plot
$\bar{g}_{1D}=g_{1D}/g$ as function of $\bar{n}$. While the plot
of $\epsilon_{long}/\epsilon_{F}$  gives a measure of the
impenetrability of the atoms, the plot of $\bar{g}_{1D}$ provides
a measure of the dimensionality, since $\bar{g}_{1D}=1$ in 1D and
decreases as the system crosses over to 3D. Since $g$ itself
depends linearly on $\omega_{0}$ the actual magnitudes of the
densities are different in the three plots shown in
Fig.~\ref{Fig1}. Along the right axis in Fig. \ref{Fig1} we plot
the scaled longitudinal energies computed from pair theory and LL
theory. We see that the density range where the pair theory
energies breaks away from the LL energies corresponds closely to
the region where $\epsilon_{long}/\epsilon_{F}$ starts to deviate
from unity. This is as we would expect since pair theory cannot
describe the Tonks regime.

\begin{figure}\vspace{-1cm}
\includegraphics*[width=\columnwidth,angle=0]{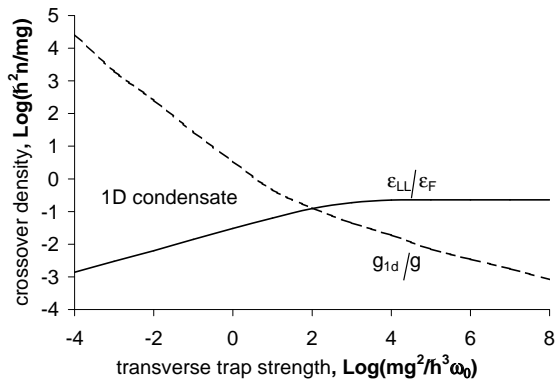}\vspace{-1cm}
\caption{Log-log plots of scaled linear densities $\hbar^{2}n/mg$
versus transverse confinement strength
$mg^{2}/(\hbar^{3}\omega_{0})$ at the BEC-Tonks crossover points
defined by $\epsilon_{long}/\epsilon_{F}=0.5$ and at the 1D-3D
crossover points defined by $g_{1D}/g=0.5$}. \label{Fig2}
\vspace{-5mm}
\end{figure}

We see that for low values of the transverse trapping potential as
in Fig.~\ref{Fig1}(a) there is a regime of density where the gas
is one dimensional but not yet a Tonks gas.  Pair theory is valid
in this region and it intrinsically allows for a ``Bose-condensed"
fraction $f$ which has the more general interpretation as the
non-trivial fraction of atoms in the ground state. Thus in this
regime we should see a 1D condensate, in the sense defined
earlier.

However, as we increase the transverse trapping potential we see
that the region where such a 1D condensate can exist gets narrower as the
curve for $\bar{g}_{1D}$ approaches the curve for
$\epsilon_{long}/\epsilon_{F}$ until the former is on the left of
the latter as we see in Fig.~\ref{Fig1}(c); this means that for
large transverse trapping potentials there will be never be a 1D condensate
and the system will pass directly from 3D to the impenetrable Tonks
gas regime. This is clearly illustrated in Fig. \ref{Fig2};  using
$\bar{g}_{1D}=0.5$ as the criterion for crossover from 1D to 3D
and $\epsilon_{long}/\epsilon_{F}=0.5$ as the criterion for
crossover to the Tonks gas, we plot the scaled densities at the
crossover points as a function of the transverse confinement
strength measured by $mg^{2}/(\hbar^{3}\omega_{0})$. The
triangular region between the two curves on the left of their
point of intersection roughly defines the range of scaled
densities and transverse trap frequencies over which a 1D condensate can
exist. On the right of the point of intersection the trap strength
is too high.

In conclusion, we have used  Lieb-Liniger theory as well as pair
theory to study the crossover of an axially homogeneous 3D Bose
condensed system to effective 1D and eventually to an impenetrable
regime. We have evaluated the densities at which the system
crosses over from 3D to 1D and then to a Tonks gas for different
transverse trap frequencies. We have demonstrated that for weak
transverse confinement there is a physically allowed regime where
a 1D axially homogeneous condensate can exist while for
sufficiently high transverse confinement the only possiblities are
a 3D condensate or a Tonks gas.

KKD thanks T. Bergeman for useful discussions. This work was
supported by Office of Naval Research grant N00014-99-1-0806 and
by the US Army Research office. \vspace{-5mm}

\begin{thebibliography}{19}
\bibitem{typo} The term $-2\lambda f$ in Eq. (16) of \cite{crossover1}
should be $-2\lambda f^2$.
\end{thebibliography}
\end{document}